\documentclass[10pt,journal,letterpaper,compsoc]{IEEEtran}

\usepackage[sort]{cite}
\usepackage{graphicx}
\usepackage{subfigure,epsfig}
\usepackage{amsmath, amsthm, amssymb}
\usepackage[all]{xy}
\usepackage{algorithmic,algorithm}
\usepackage[tight]{units}
\usepackage{nicefrac}
\usepackage{bm}
\usepackage[active]{srcltx}
\usepackage{balance}
\usepackage{url}

\begin{document}

\title{MuSA: Multivariate Sampling Algorithm for Wireless Sensor Networks}

\author{Andre L.\ L.\ Aquino,
Orlando S.\ Junior, 
Alejandro C.\ Frery, \\
\'Edler Lins de Albuquerque,
and Raquel A.\ F.\ Mini%
\IEEEcompsocitemizethanks{\IEEEcompsocthanksitem A. Aquino and A. Frery are with the Computer Institute, Federal University of Alagoas, Brazil.\protect\\
E-mail: alla@ic.ufal.br, acfrery@gmail.com
\IEEEcompsocthanksitem O. Junior and R. Mini are with the Department of Computer Science, Pontifical Catholic University of Minas Gerais, Brazil.\protect\\
E-mail: orlandosj@gmail.com, raquelmini@pucminas.br
\IEEEcompsocthanksitem E. Albuquerque is with the Department of Administration and Industrial Chemical Processes, Federal Institute of Bahia, Brazil.\protect\\
E-mail: edler@ifba.edu.br}%
\thanks{}}


\IEEEcompsoctitleabstractindextext{%
\begin{abstract}
A wireless sensor network can be used to collect and process environmental data, which is often of multivariate nature. This work proposes a multivariate sampling algorithm based on component analysis techniques in wireless sensor networks. To improve the sampling, the algorithm uses component analysis techniques to rank the data. Once ranked, the most representative data is retained. Simulation results show that our technique reduces the data keeping its representativeness. In addition, the energy consumption and delay to deliver the data on the network are reduced.
\end{abstract}

\begin{keywords}
Wireless sensor network, multivariate sampling, component analysis.
\end{keywords}}

\maketitle

\section{Introduction}
\label{sec:introducao}

\IEEEPARstart{T}{he} world around us has a variety of phenomena described by variables such as temperature, pressure, and humidity, which can be monitored by devices able of sensing, processing, and communicating. Such devices, working cooperatively, are termed wireless sensor networks (WSNs) \cite{IWYE_2002}. A characteristic that distinguishes WSNs from other networks is that nodes have very limited resources. The nodes that comprise the network are equipped with batteries and, in many applications, they will be placed in remote areas, preventing access to those elements for maintenance. In this scenario, the lifetime of the network depends on the amount of energy available to the nodes and, therefore, this limited resource must be carefully managed in order to increase the network lifetime.

The sensed data could be distinguished as univariate or multivariate. Univariate data represent samples of the same scalar phenomenon, e.g., a node that monitors only the temperature. Samples of different phenomena can be described by multivariate data. These samples are originated from different sensors in the same node. For example, a node may have sensors that monitor temperature, pressure, and humidity simultaneously.

The sampling process in WSNs can be performed in two ways, either by the sensor, or by the software after the sensing task. Direct sensor sampling, generally, cannot be reconfigured on-line and it is regular, i.e., the device is configured to take samples in regular times. Sampling by the software is more flexible, and may be tailored to meet the application requirements. For instance, when the node has not enough energy to send data, the software could reduce the amount of data to be sent to a minimum. In this paper we consider software sampling of multivariate data under energy constraints.

Based on these aspects, this work presents a \textbf{Mu}ltivariate \textbf{S}ampling \textbf{A}lgorithm (MuSA) for WSNs. MuSA uses component analysis to rank the multivariate data sensed considering only the first component scores. Based on this ranking, the sampling is performed, alleviating redundancies and maintaining the data representativeness. Two hypotheses are considered: (i)~the use of component analysis techniques to classify the multivariate data can assist the sampling maintaining its representativeness; and (ii)~the energy consumption and messages delay on the network can be reduced throughout this sampling strategy. We show evidence that these hypotheses are valid.

MuSA can be used in applications where the sensed data is multivariate.
Different component analysis techniques are available, allowing a fine tuning aiming to achieve better representativeness.
The proposal's only requirement is that the data are stationary within the observation window, but they are allowed to change their distribution among disjoint observation epochs.
Even if this assumption is not verified, the only impact on the procedure may be reducing its efficiency.

MuSA is based on a previous short conference paper by Junior et al.~\cite{OARC_2009}. 
The main improvements with respect to that publication are: 
(i)~the use of different component analysis (PCA, ICA, and robust PCA), since the previous version considered only PCA; 
(ii)~the proposition of a general sensor system model considering multivariate data; 
(iii)~the use of pseudo-real data in order to make a quantitative assessment in situations with and without departures from the underlying distribution;
(iv)~three distributions were used to simulate the data, namely, the Gaussian, the Skew Gaussian and the Student t laws; and
(v)~the network behaviour is assessed in different scenarios.
The analysis of  similar proposals with MuSA is also enhanced in this article.

This paper is organized as follows.
Section~\ref{sec:related_work} comments related work.
Section~\ref{sec:characterization} discusses multivariate sampling characterization in WSNs.
Section~\ref{sec:musa} presents the multivariate sampling algorithm (MuSA). Section~\ref{sec:representativeness} analyses simulation results regarding data representativeness.
Section~\ref{sec:network} shows how a network behaves using MuSA.
Section~\ref{sec:remarks} discuss the evaluations remarks.
Finally, Section~\ref{sec:conclusion} concludes with future research directions.

\section{Related work}\label{sec:related_work}

Adaptive sampling, which considers general models for sampling or uses algorithms that perform adaptive prediction not based on models of prior knowledge~\cite{AdLe_2003, SsKr_2006}, has been used for univariate data.
A similar approach, used for data reduction in WSNs, is the aggregation: each node decides whether or not to summarize the data, considering its energy level and the time for information delivery~\cite{BDSw_2002, JzSp_2004}.
Another technique is the data stream based reduction, in which the data is characterized as a stream of information, and specific sampling strategies are applied~\cite{alla_07, icc_07}.
Other similar strategies in WSNs are data fusion~\cite{1267073}, data compression~\cite{kimura_05} and collaborative processing~\cite{1236049}. The use of these strategies is motivated by the interest in efficient data gathering in WSNs~\cite{MMYy_2011, XYYw_2011}. 
It is important to highlight that, differently from our approach, none of these proposals considers multivariate data nor their correlation. 
In addition, these proposals are used in specific scenarios and they are not based on a general sensor system formulation.

Few techniques consider multivariate data.
Seo et al.~\cite{SJK_2005}, compare different methods:

\begin{itemize}
  \item Discrete Wavelet Transformation: it uses a hierarchical decomposition to process signals.
  \item Hierarchical Clustering: it partitions objects in groups according to their similarity.
  \item Sampling: it reduces the volume of data keeping a few samples and discarding the rest.
  \item Singular Value Decomposition: it performs a linear transformation of the data.
\end{itemize}

Regarding component analysis techniques, the most frequently employed methods to process multivariate data in WSNs are the Principal Component Analysis (PCA) and Independent Component Analysis (ICA).
Cvejic et al.~\cite{4341582} present an ICA-based algorithm to improve the fusion of surveillance images.
The method combines PCA and ICA to reduce the data.
Li and Zhang~\cite{Li_2006} propose a PCA-based algorithm to reduce multivariate data in WSNs.
The main goal is to improve transmission and management of large-size vibration sensor in structural health monitoring systems.
Roy and Vetterli~\cite{4475387} use PCA to reduce data in audio applications.
Specific applications of multivariate processing consider, for instance, satellite remote sensing images~\cite{EAAc_2009, YJJW_2008} or preprocessing data~\cite{VRDA_2011}.
Notice that our work, differently from previous ones, proposes an algorithm to perform multivariate data sampling in Wireless Sensor Networks. 
Additionally, our work allows choosing different component analysis techniques, so the algorithm can be easily tailored to obtain the desired performance according to the application.
Such fine tuning will allow sending only the most relevant data to the sink.

\section{Multivariate sampling characterization in WSNs}\label{sec:characterization}

\begin{figure}[h]
\begin{displaymath}
 \xymatrix{\mathcal{N} \mid E \ar[r]^P & \mathbf{V} \ar[r]^{\mathbf{S}(h,k)} \ar[d]^{R} & \mathbf{V}' \ar[r]^{\Psi} & \mathbf{V}'' \ar[r]^{\widehat{P},h} & \mathbf{V} \ar[d]^{\widehat{R}}\\
            & D & & & \widehat{D}}
\end{displaymath}
\caption{Representation of a WSN system.}
\label{fig:wsn_appl_behavior}
\end{figure}
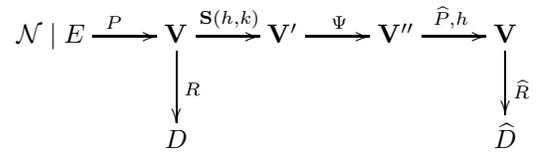

A general sensor system can be modelled according to the diagram presented in Fig.~\ref{fig:wsn_appl_behavior}.

In this diagram, which is based on the one presented by Frery et al.~\cite{DataDrivenPerformanceEvaluationWirelessSensorNetworks}, $\mathcal N$ represents the environment and the process to be measured.
The study is restricted to (denoted as ``$\mid$'') $E$, the time-space domain and topological characteristics of the monitored area.
The phenomenon of interest is $P$, and $\mathbf{V}$ is its domain, i.e., $\mathbf V$ is the set of all possible phenomena.
An example of this model is a forest ($\mathcal N$), with our attention restricted to a critical area $E$ where the occurrence of fire is not acceptable.
The phenomenon of interest could be the pair ``temperature-humidity'', with infinite precision in space, time and the measures.

If noise-free observations were available without loss of information, a set of ideal rules ($R$) leading to ideal decisions ($D$) would be devised.
Fully dependable and precise measures of temperature and humidity would lead to such decision rules, for instance when to send forest rangers.
Due to the complexity of the systems in which the WSNs operate, this is  unachievable.

The following sections present the more realistic model we will discuss.

\subsection{Sensory behaviour}

Instead of the ideal situation, a set of $o$ observer nodes, $\mathbf{S} = (S_1, \ldots, S_o)$ is deployed to perform a sampling over $\mathbf{V}$.
Each node $S_i$ is aware of its position ($h_i$), and the operations it can perform are described by a characteristic function $k_i$.
In Fig.~\ref{fig:wsn_appl_behavior}, $h$ denotes the collection of all positions, and $k$ denotes the set of all characteristic functions.

One of such operations is sensing.
If the phenomenon of interest can be described by the function $f$, say humidity (in \unitfrac{mg}{m$^3$}) and temperature (in Celsius degrees) at instant $t$, i.e., $f(t)=(f_1,f_2)(t)$, a node will record instantaneous values proportional to the integral of $f(t)$ within a nearby area.
Another operation the node can perform is the local computation of the \AA{}ngstrom index~\cite{SMHT_2004} which is a simple fire-danger rating system given by
$$
f_3(t) = \frac{f_1(t)}{20} + \frac{27-f_2(t)}{10}.
$$
With this, the data available to node $S_i$ depends on its location $h_i$, and can be denoted as
$$
\bm s_i(f(t)) = k_i(f_1,f_2)(t) =  \bigl(f_1(h_i),  f_2(h_i), f_3(h_i)\bigr)(t).
$$
Each component represents, respectively, the ideal (point wise) temperature, the ideal humidity, and the ideal \AA ngstrom index at instant $t$ as recorded in $h_i$.
Node localization is a important issue in wireless sensor networks~\cite{ZyYl_2012}. 
However, its study is outside the scope of this work. 
We  consider that the best solution for node localization could be used without compromising the multivariate reduction performed. 
This occurs because all nodes will execute the same localization algorithm, so the energy consumption and processing will increase proportionally in all nodes.

Another node $S_j$ that monitors the same quantities could have its operation characterized by
\begin{eqnarray*}
\lefteqn{\bm s_j(f(t)) = k_j(f_1,f_2)(t)} \\
& = & \Biggl(\int_{d_{1,j}} f_1 , \int_{d_{2,j}} f_2 , \frac{\int_{d_{1,j}} f_1}{20} + \frac{27 - \int_{d_{2,j}} f_2}{10} \Biggl)(t),
\label{eq:general_sj}
\end{eqnarray*}
where the domains of integration are $d_{1,j}\in\mathbb R$ and $d_{2,j}\in\mathbb R_+$.
Differently from the previous sensor, this last one performs records of the resulting values which are found by integrating the signal in a nearby region of capturing. 
Notice that different domains of integration are considered for each the temperature and the humidity, namely $d_{1,j}$ and $d_{2,j}$.
Thus, more realistic measures of the temperature, humidity, and \AA ngstrom index where found.

Assuming that all nodes record these three values, namely humidity, temperature, and \AA ngstrom index, in a synchronized fashion, the collection of $o$ nodes samples the phenomenon $f$ at $o$ positions, and each position yields a multivariate observation.
Each observation belongs to $\mathbb R_+\times \mathbb R\times \mathbb R$.

The collection of all multivariate observations is of the form $(\bm s_1,\dots,\bm s_o)(t)$, where each $\bm s_j$ is given by previous equation, so the network records a point in $(\mathbb R_+\times \mathbb R\times \mathbb R)^o$ at each instant.

If $n$ instants are recorded, namely, $t_1,\dots,t_n$, the information captured by the network is
$$
\left(
\begin{array}{c}
(\bm s_1,\dots,\bm s_o)(t_1)\\
(\bm s_1,\dots,\bm s_o)(t_2)\\
\vdots\\
(\bm s_1,\dots,\bm s_o)(t_n)\\
\end{array}
\right),
$$
which is a $3\times o\times n$-fold real-valued vector.
We assume that the data is stationary in each temporal window of size $n$.
The technique we propose allows changes in distribution (mean, variance and covariance structure etc.) among windows, since all procedures are built from scratch for each $n$-tuple dataset.

This is a highly redundant set of data with values in $\mathbf V'= \big((\mathbb R_+\times \mathbb R\times \mathbb R)^o\big)^n$.
The redundancy is spatial (nearby observations are likely similar), temporal (measures in $t_i$ and $t_{i+1}$ are likely similar) and local (in our example, the triplet in each node can be reduced to a pair of values without loss of information).
Data reduction is, therefore, both feasible and desirable in order to prevent power depletion and, with it, the death of the network.

More generally, instead of three, we will consider $p$ real variables available in each node, having, thus, a $p\times o\times n$ data set.

\subsection{Data reduction}

Using the whole $\mathbf{V}'$ may be infeasible, so some data reduction should be applied.
As discussed earlier, sending large amounts of data can be very costly in terms of energy and bandwidth.
Besides that, messages delivery time may suffer from excessive delay, rapidly degrading the network lifetime.

We propose sampling techniques for reducing the delay and energy consumption.
As outlined in Fig.~\ref{fig:wsn_appl_behavior}, multivariate data sampling strategy is a transformation of the form
$$
\Psi \colon \mathbb{R}^{p \times o \times n} \rightarrow \mathbb{R}^{p\times o \times n'},
$$
where $n' < n$ is the number of samples over $\mathbf{V}'$, so we keep the number of variables $p$ and the number of nodes $o$, but reduce the number of observations.

For the sake of simplicity, we will describe this sampling strategy nodewise, i.e., the whole transformation $\Psi$ is the result of applying operations on each node $1\leq i\leq o$: $\Psi=(\Psi_1,\dots,\Psi_o)$.
Each nodewise transformation produces a transformation $\Psi_i\colon \mathbb{R}^{p \times n} \rightarrow \mathbb{R}^{p\times n'}$.

In our proposal, the sampling $\Psi_i$ is the composition of three functions, namely, a components transformation ($\Psi_C$), followed by a ranking ($\Psi_O$) and then the sampling ($\Psi_A$):
$$
\Psi_i = \psi_A \circ \psi_O \circ \psi_C.
$$

Three components transformations will be considered: principal components, robust principal components and independent component analysis.
The ranking we will employ consists in choosing the most important component.
The sampling will consist in choosing the original observation indexed by the result of the previous stage.
In the following we will formalize these steps.

The components transformation ($\psi_C$) denotes any operation on the data space yielding the same number of dimensions.
\begin{eqnarray*}
 C_{n,p}= \psi_C(\bm s_{i}) 
& = & 
\psi_C
\left( \begin{array}{c}
( s_{i1}, \dots, s_{ip}) (t_1)\\
\vdots\\
( s_{i1}, \dots, s_{ip}) (t_n)\\
\end{array}
\right) \\ 
& = &
\left( \begin{array}{ccc}
C_{1\,1} & \cdots & C_{1\,p} \\
\vdots &  & \vdots \\
C_{n\,1} & \cdots & C_{n\,p} 
\end{array} 
\right) \\ 
& = &
\left(
\begin{array}{c} C_1 \\ \vdots \\ C_n \end{array}\right)
,
\end{eqnarray*}
where $\bm s_i$ is the collection of observations gathered by node $i$ along $t_1,\dots,t_n$.

Different $\psi_C$ transformations can be applied as, for instance,

\begin{itemize}
  \item \textbf{Principal component analysis (PCA)}: It is a linear transformation in the data set which produces a new uncorrelated data set~\cite{Krzanowski1988, Svan_19987} which can be conveniently reduced with little loss of information, due to the way the new data are formed.
Considering $\bm s_i \in \mathbf{V_i}'$ as input data, calculate $C_j = [\bm s_i - \overline{\bm s_i}] E$, where $E$ is the matrix whose columns are the eigenvectors of the (possibly standardized) covariance matrix of the data $\bm s_i$.

  \item \textbf{Robust principal component analysis (robust-PCA)}: This class of techniques was proposed as a means to tackle possible shortcomings of traditional PCA in the presence of outliers. The main difference is the way the eigenvalues are obtained: each element of the covariance matrix is computed robustly~\cite{Ruym_1981, HRVb_2005}, or, alternatively, robust eigenvalues satisfying certain properties are sought~\cite{Maronna:PCA}.

  \item \textbf{Independent component analysis (ICA)}: It also performs a linear transformation in the data, looking for statistically independent new components. It can reduce, increase or maintain the original data dimensions~\cite{195312, 802359}. Among the many alternatives available, we applied a nonlinear PCA algorithm based on approximations to neg-entropy which is more robust than kurtosis based measures and fast to compute as provided by R~\cite{RManual}.
\end{itemize}

To perform the ranking ($\psi_O$), consider $\Delta$ the operator that sorts the indexes of the first column of $C$ in descending order:
$$
\Delta \left(\begin{array}{c} C_{11} \\ \vdots \\ C_{n1} \end{array}\right) = \Delta(1, 2, \ldots, n) = (\delta_1, \delta_2, \ldots, \delta_n),
$$
such that $C_{\delta_1, 1} \leq C_{\delta_2, 1} \leq C_{\delta_n, 1}$.
Therefore, $\psi_O$ consists of sorting the first component values in $C$.

\begin{figure}[h]
    \centering
    \includegraphics[width=.4\textwidth]{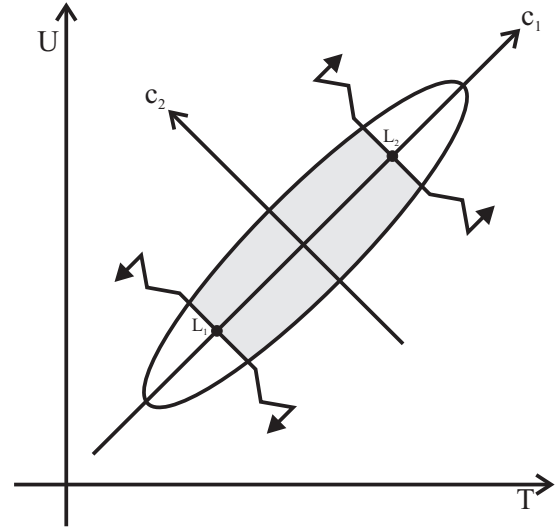}
	\caption{Determination of sample: $T$ and $U$ original variables, $C_1,C_2$ principal components, $L_1,L_2$ lower and upper bounds in the transformed space.}	
	\label{fig:sampling}
\end{figure}

Finally, the sampling process ($\psi_A$) consists of choosing the elements in $\bm s_i$ whose ranks in $\Delta$ are neither the smallest nor the biggest.
These $n'<n$ (assumed even for simplicity) elements are defined by the observations with indexes $(\delta_{\frac{n'}{2}}, \ldots,  \delta_{n-\frac{n'}{2}})$.
Denote such sample $\bm s_i'$.
This procedure is illustrated in Fig.~\ref{fig:sampling}, where the gray area contains the indexes to be sampled.

This strategy differs, in general, from criteria based on the Euclidean and on the Mahalanobis distances to the center of mass, which form the core of the k-means strategy.
The set of possible values of the sampled data is $\mathbf{V''_i}\subset \mathbf{V'_i}$.

\subsection{Reconstruction, measures of quality, and decisions}
\label{subsec:rules}

After sampling, some kind of reconstruction must be performed by the sink node in order to take decisions about the phenomena under study.
In Fig.~\ref{fig:wsn_appl_behavior} the reconstruction is represented by $\widehat{P}$ that uses the whole network localization information, represented by $h = \{ h_i : 1\leq i\leq o\}$, where $h_i$ is the position of sensor $S_i$ in $\mathbb R^2$ or $\mathbb R^3$.

New rules over $\mathbf{V}$ are represented by $\widehat{R}$ and they lead to the set of decisions $\widehat{D}$.

Considering the need to perform the sampling $\Psi$ over $\mathbf{V}'$, a fundamental aspect to be analysed is the impact of this processing on the decisions.
In this work, following Frery et al.~\cite{DataDrivenPerformanceEvaluationWirelessSensorNetworks}, we assess this impact analysing the sampled data with respect to the original one, rather than checking changes in decisions.
If ideal reconstruction techniques were available, the decisions would be the same.
Such assessment is performed with the following measures:

\begin{itemize}
\item Analysis of variance -- ANOVA~\cite{anova} aims to evaluate whether there are significant differences between the averages of the original data set and the reduced one. The test statistic is
    $$
    T = \lambda^{2}_{B}/\lambda^{2}_{W},
    $$
    where $\lambda^{2}_{B}$ is the variance between sets $\bm s_i$ and $\bm s_i'$ and ${\lambda^{2}_{W}}$ the variance within sets.
Based on this calculation, the \emph{p-value} is used to determine if the null hypothesis $H_{0}$ must be rejected.
In this case, the null hypothesis models that there are no significant differences between the variances of the two sets.
Values of \emph{p-value} below 0.05 provide sufficient evidence to reject the null hypothesis at the $95\%$ confidence level. By convention, $\widehat{R}_{\text{ANOVA}}$ will be used to indicate the use of this test.

\item Our measure of error is the maximum relative absolute error between averages of original data $\bm s_i$ and the reduced one:
    $$
    \widehat{R}_{ERROR} = 100 \max_j
\frac{
|\overline{s}_{ij}(\cdot) - \overline{s}'_{ij}(\cdot)|}{ |\overline{s}_{ij}(\cdot)|
}.
    $$
\end{itemize}

\section{Multivariate Sampling Algorithm}
\label{sec:musa}

Based on the characterization presented in previous section, here, it is presented MuSA -- \textbf{Mu}ltivariate \textbf{S}ampling \textbf{A}lgorithm, that implements the $\Psi$ processing.
The goals of MuSA are: to allow the use of different component analysis techniques; and to diminish redundancies and minor details, getting a subset of the original data with minimal information loss.

To reduce $\bm s_i$, MuSA performs in four steps: step 1, the original set of sensory data $\bm s_i$ is used to calculate the components ($\psi_C$); step 2, the first component $C_{\,\cdot\, 1}$ is ranked ($\psi_O$); step 3, the scores $\delta_1, \ldots, \delta_n$ are used to determine the lines in $\bm s_i$ that will compose the reduced data $\bm s_i'$. These scores are the intermediate indexes, i.e., the values that represent the samples with similar variance ($\psi_A$); and step 4, the reduced data set $\bm s_i'$, containing the most relevant data, is obtained.
A pseudo-code of MuSA is given in Algorithm~\ref{alg:pca}.

\begin{algorithm}
\caption{Multivariate sampling algorithm ($\Psi$ processing)}
\label{alg:pca}
\begin{algorithmic}[1]
    \REQUIRE $\bm s_i$ -- original data, $n'$ -- reduction size
    \ENSURE $\bm s_i'$ -- reduced data\\*[.05in]
    \STATE $C_{n,p} \leftarrow \psi_C(\bm s_i)$ \COMMENT{Component calculation} \label{line_1}
    \STATE $\delta \leftarrow \Delta(C_{\,\cdot\, 1})$ \COMMENT{Sort indexes} \label{line_2}
    \STATE $L \gets (\delta_{\frac{n'}{2}}, \ldots,  \delta_{n -\frac{n'}{2}})$ \COMMENT{Get the indexes	 used in sampling} \label{line_3}
    \FOR {$j\leftarrow 1$ \textbf{to} $n'$} \label{line_4}
        \STATE $\bm s_{i_{j \,\cdot}}' \leftarrow \bm s_{i_{L_j \,\cdot}}$ \COMMENT{Sampling action} \label{line_5}
    \ENDFOR \label{line_6}
    \STATE $\bm s_i' \leftarrow Sort(\bm s_i')$ \COMMENT{Sort the final sample considering the original order of arrival, `optional step'} \label{line_7}
\end{algorithmic}
\end{algorithm}

A line-by-line of Algorithm~\ref{alg:pca} is the following:

\begin{itemize}
\item In Line~\ref{line_1}, we have the calculation of the components through the chosen technique.
The complexity of this PCA calculation is $O(p^{2}n)$~\cite{bb42143}, where $n$ represents the number of samples and $p$ the number of variables.
The complexity order of computing the first robust-PCA component is $O(pn)$~\cite{937541}.
For the calculation of ICA, considering the FastICA algorithm, the order can be estimated in $O(pn)$~\cite{fastICA_06}.

\item In Line~\ref{line_2}, the first component $(C_{11}, \dots, C_{n1})$ is sorted. Its complexity, considering a simple quicksort algorithm, is $O(n\log_{2}{n})$.

\item In Line~\ref{line_3}, we discard the extreme values, which is an $O(1)$ operation.

\item In Lines \ref{line_4} -- \ref{line_6}, we build the reduced output data, with complexity order is $O(n')$.

\item In Line~\ref{line_7}, the sampling $\bm s_i'$ is sorted. Its complexity, considering a simple quicksort algorithm, is $O(n'\log_{2}{n'})$, since only the lines are sorted.
\end{itemize}

Thus, total time complexity using PCA is
$$
O(p^{2}n) + O(n\log_{2}{n}) + O(n') + O(n'\log_{2}{n'}) = O(p^{2}n).
$$
Considering ICA or robust-PCA, total time complexity is
$$
O(pn) + O(n\log_{2}{n}) + O(n') + O(n'\log_{2}{n'}) = O(pn).
$$

For the space complexity, consider the matrices $\bm s_i$, $\bm s_i'$, $C$, the mean vector and the (possibly scaled) matrix of eigenvectors used in PCA.
Space complexity is thus given by
$$
3O(p n) + O(p) + O(pn')= O(pn).
$$

Since each source node sends $\bm s_i'$ to sink, communication complexity is $$O(pn'\#\mathrm{hops}),$$ where `$\#\mathrm{hops}$' is the largest route in the network.

\section{Data representativeness}
\label{sec:representativeness}

In this section, we evaluate the data representativeness.
The goal is to determine if the reconstructed data $\mathbf{V}$ represent satisfactorily the environment monitored. In this way, the analyses applied are $\widehat{R}_{ANOVA}$ and $\widehat{R}_{ERROR}$ defined in Section~\ref{subsec:rules}.

\subsection{Methodology}
\label{subsec:methodology}

We use pseudo-real data in order to make a quantitative assessment in situations with and without departures from the underlying distribution.

The real data are 19-dimensional observations of environmental phenomena, among them the concentration of pollutants \emph{n-hexane}, \emph{methylcyclopentane}, \emph{toluene}, \emph{p-xylene}, and \emph{1,3,5-TMB}.
The sensed data $\bm s_i$ consists of 72 of these observations and, as reported in \cite{edler_2007}, each data is the result of four hours of measurements.
The individual measurements that led to the observations are not available and, thus, will be simulated.
The simulation will be made retaining the mean and covariance structures, but assuming three cases:  Gaussian, Skew Gaussian and heavy-tailed Student t distributions.
We will call these simulated data ``pseudo-real" data.

\begin{figure}[h]
   \centering
   \subfigure[Linear scale]{
       \includegraphics[width=.45\textwidth]{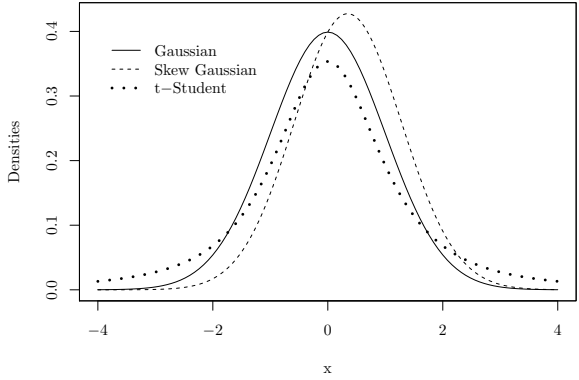}
       \label{subfig:Density}
   }
\\
   \subfigure[Semilogarithm scale]{
       \includegraphics[width=.45\textwidth]{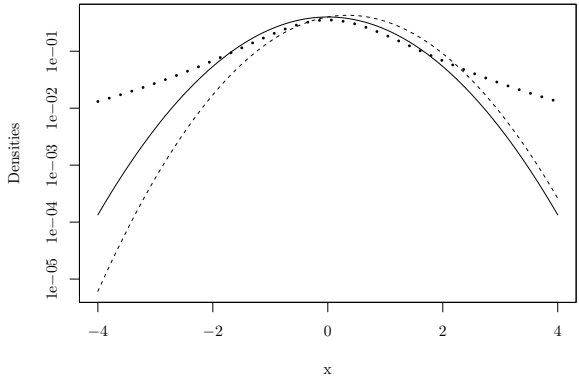}
       \label{subfig:DensityLog}
   }
   \caption{Densities of distributions}
   \label{fig:density}
\end{figure}

Figs.~\ref{subfig:Density}-\ref{subfig:DensityLog} present the densities which characterize the Gaussian, the Skew Gaussian and the Student t distributions with zero mean in linear and semilogarithmic scales.
The Skew Gaussian law here depicted has asymmetry parameter $1/2$, while the degrees of freedom of the Student t distribution is $2$.
The former describes situations where there is skewness, while the latter models heavy-tailed situations.

All evaluations are made through simulations, with algorithms implemented in the $R$ platform~\cite{RManual}, whose excellent numerical properties were assessed by Almiron et al~\cite{AlmironSilvaMM:2009}.
The number of necessary replications was calculated, following Jain~\cite{jain_91}, as
$$
rounds = \Big{(} \frac{100\, c \widehat{\sigma}}{ x\, \overline{error}} \Big{)}^{2},
$$
where $c=1.96$, $\widehat{\sigma}$ is the sample standard deviation found in the five first replications of the worst situation, $\overline{error}$ is the average value and $x$ is the percentage of the average that we want to get as deviation, i.e., the precision, that in this case was $5\%$.
Each scenario was, thus, executed with $1000$ independent data sets.

In order to explore the level of reduction supported by the applications, so that sampling will not compromise the decisions $\widehat{D}$, we use two levels of reduction to compose $\bm s_i'$, $n' = n/2$ and $n' = \log_{2}{n}$.
The former is usually acceptable in practice, while the latter can be considered a \textit{hardcore} data reduction only intended for providing a reference and not intended to be used in practice.
Applications have different needs and characteristics, so different levels of reduction between these two extremes can be applied.

Our data representativeness evaluation considers the PCA, robust-PCA, and ICA components analysis techniques.

The number of pseudo data generated in each interval varies in $\{10, 20, 30, 40, 50\}$, i.e. the reading $n$ is $\{720, 1440, 2160, 2880, 3600\}$. In data representativeness we do not consider a specific network topology, because the main objective is show the fidelity of the data reduced when compared to the sensed one.

\subsection{Analysis of variance -- $\widehat{R}_{ANOVA}$}
\label{subsec:variance}

Table~\ref{tab:anova} presents the results considering the robust-PCA.
As expected, the $n/2$ data reduction leads to smaller differences between observed and sampled data than the $\log_2 n$ strategy; in fact, the former produces data which is indistinguishable from the original information.
It is also not surprising that Gaussian data are best dealt with by the reduction, while the worst scenario is provided by the Skew Gaussian model.

\begin{table*}[t]
\begin{center}
\caption{Analysis of variance (\emph{p-value})}
\label{tab:anova}
\begin{tabular}{ccccccccccc}
\hline
    Evaluated
    & \multicolumn{2}{c}{{\scriptsize \textbf{($n$ = 720)}}}
    & \multicolumn{2}{c}{{\scriptsize \textbf{($n$ = 1440)}}}
    & \multicolumn{2}{c}{{\scriptsize \textbf{($n$ = 2160)}}}
    & \multicolumn{2}{c}{{\scriptsize \textbf{($n$ = 2880)}}}
    & \multicolumn{2}{c}{{\scriptsize \textbf{($n$ = 3600)}}}
    \\
    distribution
    & $n/2$ & $\log_{2}{n}$
    & $n/2$ & $\log_{2}{n}$
    & $n/2$ & $\log_{2}{n}$
    & $n/2$ & $\log_{2}{n}$
    & $n/2$ & $\log_{2}{n}$
    \\
\hline\hline
    Gaussian      & 0.0 & 6.21e-09 & 0.0 & 1.59e-09 & 0.0 & 9.49e-11 & 0.0 & 1.21e-10 & 0.0 & 9.89e-11\\
    Skew Gaussian & 0.0 & 3.72e-03 & 0.0 & 9.73e-04 & 0.0 & 4.80e-05 & 0.0 & 4.27e-05 & 0.0 & 1.13e-04\\
    Student t     & 0.0 & 5.88e-04 & 0.0 & 5.19e-10 & 0.0 & 1.55e-12 & 0.0 & 3.33e-08 & 0.0 & 1.40e-07\\
\hline
\end{tabular}
\end{center}
\end{table*}

The results indicate that the proposed sampling does not introduce identifiable distortions in the original data.
The PCA and ICA results provide even stronger support for not rejecting the hypothesis of same distribution.

The reduced data set $\bm s_i'$  could, thus, represents  the original data set $\bm s_i$ satisfactorily.
Therefore, decisions $\widehat{D}$ taken based on $\widehat{R}_{ANOVA}$ are the same using the original and the reduced data.

\subsection{Analysis of relative absolute error -- $\widehat{R}_{ERROR}$}
\label{subsec:erro}

\begin{figure}[h]
   \centering
   \includegraphics[width=.45\textwidth]{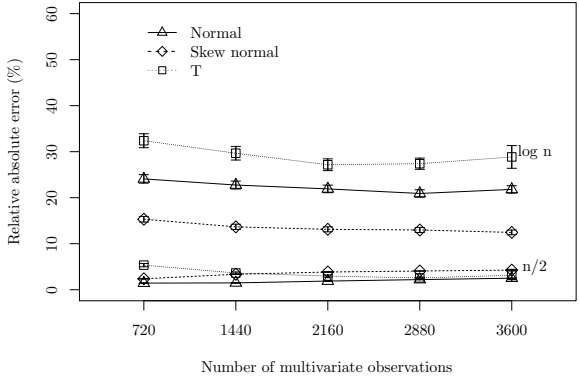}
   \caption{Analyses of MuSA - PCA results}
   \label{fig:AVG_ERRO_PCA}
\end{figure}

\begin{figure}[h]
   \centering
   \includegraphics[width=.45\textwidth]{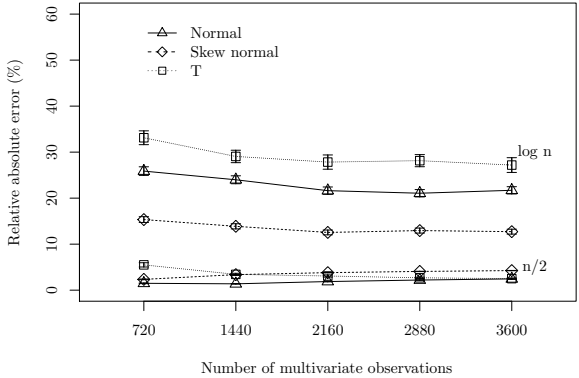}
   \caption{Analyses of MuSA - ICA results}
   \label{fig:AVG_ERRO_ICA}
\end{figure}

\begin{figure}[h]
   \centering
   \includegraphics[width=.45\textwidth]{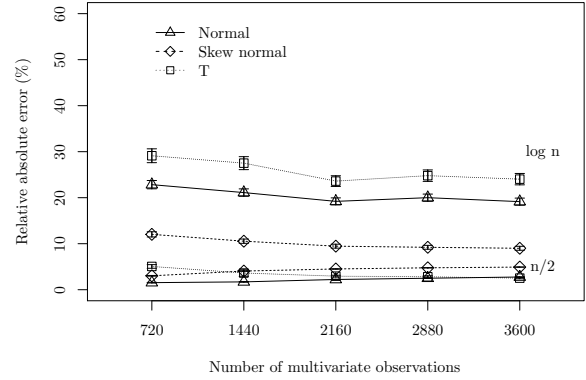}
   \caption{Analyses of MuSA - PCA-robust results}
   \label{fig:AVG_ERRO_PCA_R}
\end{figure}

The second analysis considers $\widehat{R}_{ERROR}$, the absolute relative error.
The mean value and $95\%$ asymptotic symmetric confidence intervals are shown in Figs.~\ref{fig:AVG_ERRO_PCA} -- \ref{fig:AVG_ERRO_PCA_R}.
The error of strategies that retain $\log_2 n$ of the data is always above than those than employ $n/2$ reduction.

Considering the  $n/2$ sampling strategy, the results obtained with the three techniques were very satisfactory and are practically equal in terms of relative absolute error: close to $5\%$.
The $R_{ERROR}$ decreases whenever the amount of sensed data $\bm s_i$ increases, regardless the technique.
This occurs because a higher amount of data is generated for each phenomenon.

Considering the sampling $\log_{2}{n}$, the worst result is close to $30\%$, and it is consistently observed under the Student t distribution regardless the technique.
This occurs because this distribution is more sparse, i.e., heavy-tailed.
The best results, which are consistently close to $10\%$, are observed under the Skew Gaussian model, regardless the technique.
This is probably due to the fact that the distribution is more concentrated and, then, the MuSa data retains are excellent representatives of the whole sample.

Albeit the differences are small among techniques, the smallest errors are consistently produced by robust-PCA, while the biggest ones are related to ICA, regardless the sample size.
This occurs, according to Hyv\"arinen~\cite{802359}, because the first independent component may not have the highest percentage of data variance, making the sampling process less efficient.

The scalability of the proposed algorithm in terms of the amount of sensored data is confirmed, since with all used techniques, when we increase $\bm s_i$, $\widehat{R}_{ERROR}$ is diminished or kept practically the same.

\section{Network behaviour evaluation}
\label{sec:network}

This section employs network behaviour parameters as a means for evaluating the MuSA algorithm.

Since communication is the task that consumes most energy in WSNs, reducing the amount of transmitted data will reduce the amount of consumed energy.
Henceforth, network behaviour will be evaluated to show the benefits of multivariate sampling in terms of energy consumption and delay to deliver data to sink.

The simulations are performed using the bulk of data $\bm s_i'$.
Sampling does not affect the network performance since the amount of energy required for processing is negligible when compared with that of transmission.

\subsection{Methodology}
\label{subsec:methodology_net}

Network evaluation is made through Network Simulator~2 (NS-2) version~2.33 \cite{ns2}. 
The number of necessary simulations was calculated in the same fashion as above, yielding $30$ topologies.
We consider a flat network with a shortest path tree routing and nodes with the same hardware.
The network density is kept constant and the source nodes are uniformly and independently distributed (random deployment) in the sensory area.
The parameters used in the simulations are presented in Table~\ref{tab:parametros}.

\begin{table}[h]
\begin{center}
\caption{Network simulation parameters} \label{tab:parametros} \centering
\begin{tabular}{p{3.5cm} p{3.5cm}}
\hline
\textbf{Parameter}       & \textbf{Values} \\
\hline\hline
Network size             & Varies with density \\
Sink location            & $0$, $0$            \\
Simulation time (s)      & $1100$              \\
Radio range (m)          & $50$                \\
Bandwidth (kbps)         & $250$               \\
Source location          & Random              \\
Traffic start (s)      & $500$                 \\
Traffic end (s)          & $600$                 \\
Data rate (s)            & $60$                  \\
Initial energy (J)       & $100$                 \\
\hline
\end{tabular}
\end{center}
\end{table}

In order to evaluate only the performance of sampling, the trees are built just once before the traffic starts.
Considering the main objective of this evaluation, namely network behavior with sampling, rebuild the tree could interfere the assessment. 
However, if the tree has to be rebuilt all nodes will execute the same algorithm, so the energy consumption and processing will increase proportionally in all nodes.
In addition, sampling is performed after the data is sensed and routed to the sink. 
Currently, we do not consider the use of new sampling during routing~\cite{AaEn_2009} in different nodes. 
Therefore, the use of different routing strategies will not affect the MuSA data quality neither the network behaviour.

The network size varies but the network density is kept constant in $8.48$ and is obtained through $net_{t} = \sqrt{\pi a_{r}^{2} |S|/8.4791},$ where $a_{r}$ is the radio range and $S$ the number of nodes.
Package discarding by limited queue size is avoided.
The radio range and bandwidth consider the specification of the MicaZ node~\cite{xbow}.
The simulation time was set to \unit[$1100$]{s}, where the first \unit[$500$]{s} are used for building and configuring the network; the final \unit[$500$]{s} are used to allow the remaining packages in the network be transmitted.
Thus, the actual data traffic on the network employs \unit[$100$]{s}.
Moreover, the initial energy used was \unit[$100$]{J}, so that the network nodes never had their energy depleted.

We consider the following two metrics: energy and packet delay, with varying data size, network size, and number of source nodes.
We evaluate $\bm s_i'$ with two reduction strategies, namely $n' = \{n/2, \log_{2}{n}\}$.

\subsection{Energy evaluation}
\label{subsec:energy}

\begin{figure}[h]
   \centering
   \subfigure[Varying data size]{
       \includegraphics[width=.45\textwidth]{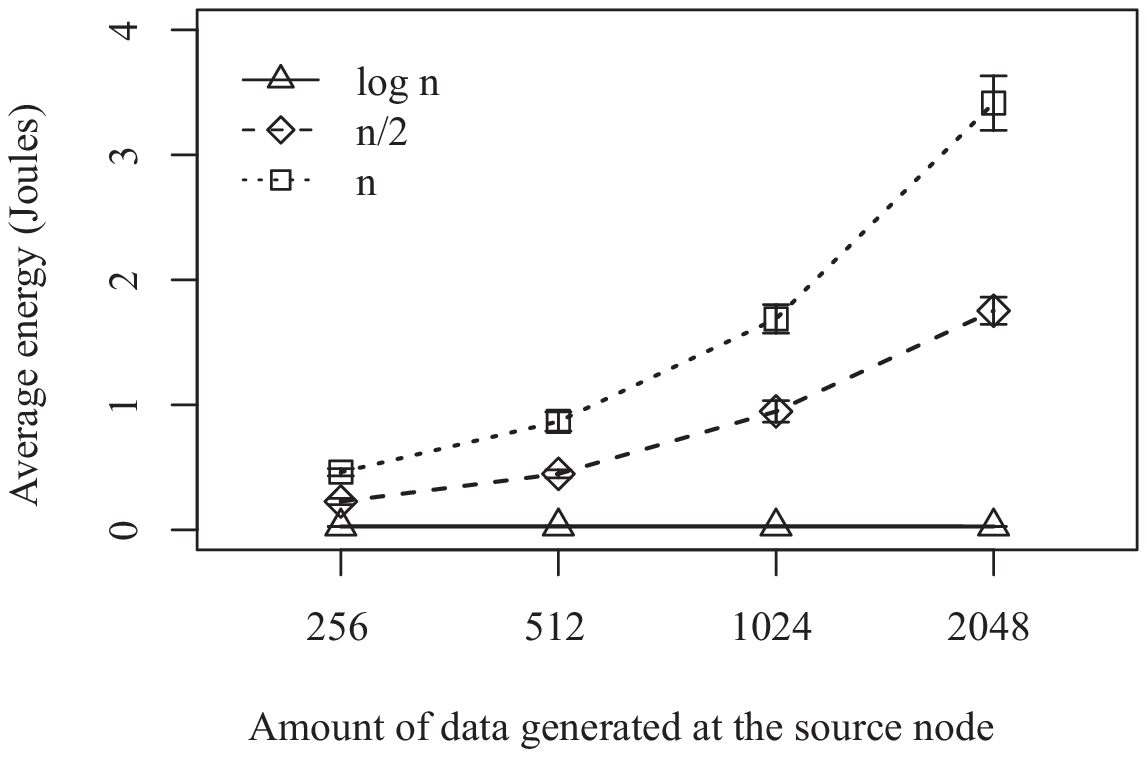}
       \label{subfig:ogkmv_STREAM_SIZE_energy_5fontes}
   }\\
   \subfigure[Varying number of nodes on the network]{
       \includegraphics[width=.45\textwidth]{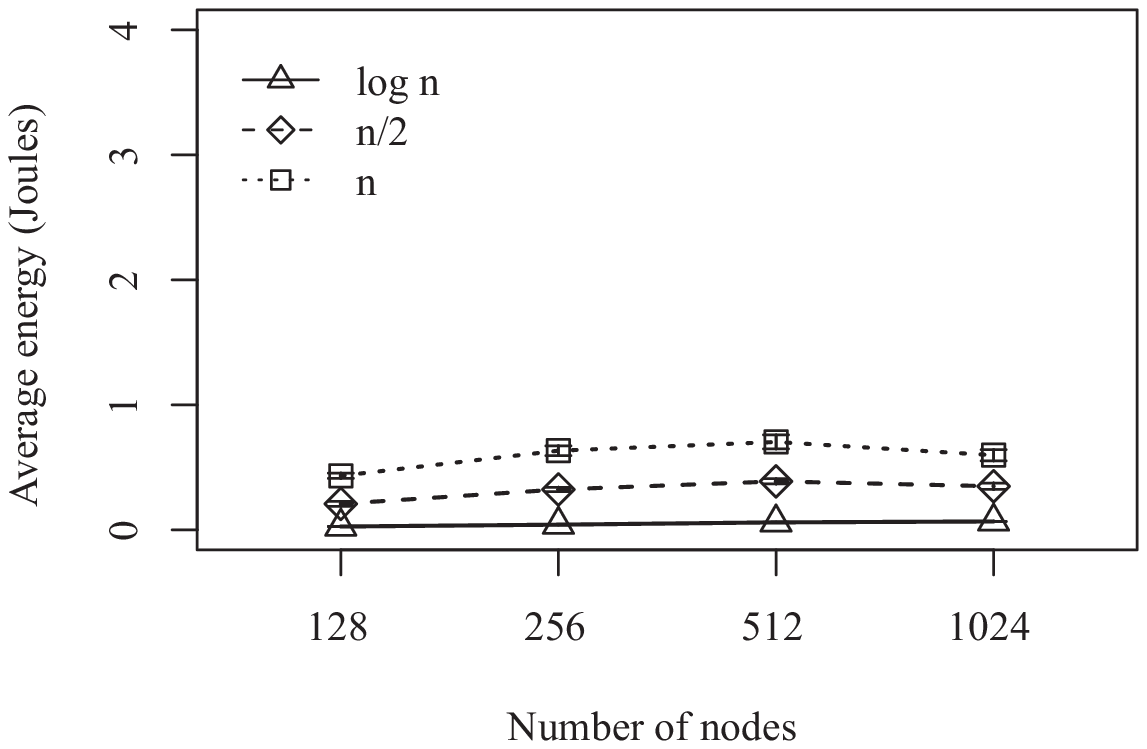}
       \label{subfig:ogkmv_NUM_NODES_energy_5fontes}
   }\\
   \subfigure[Varying number of source nodes]{
       \includegraphics[width=.45\textwidth]{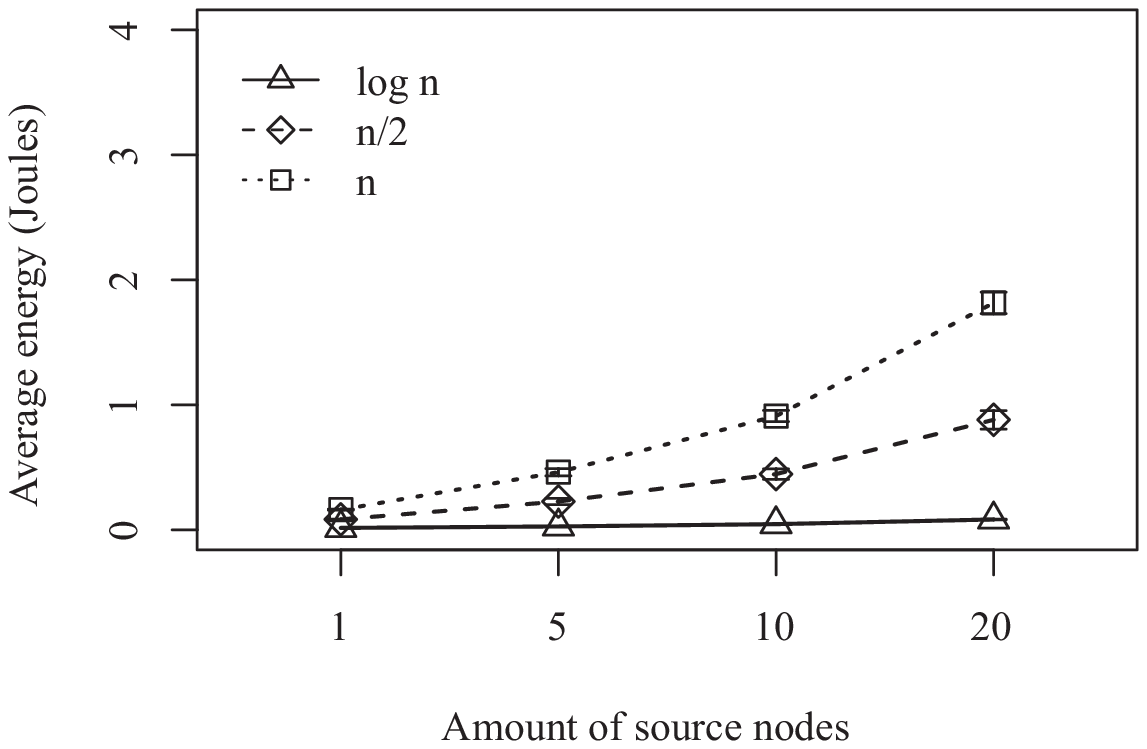}
       \label{subfig:ogkmv_SOURCE_energy}
   }
   \caption{Evaluation of energy consumption}
   \label{fig:ogkmv_sensoriamento_energy}
\end{figure}

The first analysis considers the energy consumption. 
Data transmission for offline storage, query and data analysis in a central node is an expensive procedure in WSNs, since wireless communication consumes a large amount of energy~\cite{QMWy_2006, WRNM_2011}. 
The energy consumption of multivariate data processing can be assessed by computing the algorithm time complexity; this is done in Section~\ref{sec:musa}. 
However, the energy consumption of data processing is irrelevant when compared to data transmission. 
Therefore, in our energy evaluation we only consider the data transmission energy consumption.
Simulation results are presented in Fig.~\ref{fig:ogkmv_sensoriamento_energy}.

Fig.~\ref{subfig:ogkmv_STREAM_SIZE_energy_5fontes} shows the dependence of energy on the size of $\bm s_i$, which varied in $n = \{256, 512, 1024, 2048\}$.
The number of source nodes was fixed in $5$, and the amount of nodes in the network in $128$.
It is possible to observe that energy consumption increases significantly when we increase the data size, which occurs because the traffic inserted on the network also increases.
However, with a sampling $\log_{2}{n}$ this variation is not observed, since the amount of traffic on the network is very small.

The MuSa behaviour when the number of nodes in the network varies in $\{128, 256, 512,1024\}$  is presented in Fig.~\ref{subfig:ogkmv_NUM_NODES_energy_5fontes}, for a fixed amount of data generated $n = 256$ and $5$ source nodes.
It can be seen that the energy consumption is almost constant as the number of nodes in the network increases, since the network traffic is also practically the same.
However, a small increase in energy consumption can be observed when sending $n$ data, as the number of nodes in the network increases; this is because more nodes are required to forward the packages to the sink.

Finally, we evaluated the energy consumption varying the number of source nodes, shown in Fig.~\ref{subfig:ogkmv_SOURCE_energy}.
For this, we varied the number of source nodes in $\{1, 5, 10,20\}$, keeping constant the amount of data in $n = 256$ and the number of nodes in $128$.
Energy consumption increases considerably with the number of source nodes.
This occurs because the amount of traffic on the network also increases considerably with the number of sources.
Once more, it is noticeable that energy consumption diminishes as the amount of transmitted data is reduced.
Moreover, again, it can be seen that when $\log_{2}{n}$ sampling is applied, there is almost no increase in the energy consumption, since the amount of data sent on the network is kept very small.

\subsection{Packet delay evaluation}
\label{subsec:delay}

\begin{figure}[h]
   \centering
   \subfigure[Varying data size]{
       \includegraphics[width=.45\textwidth]{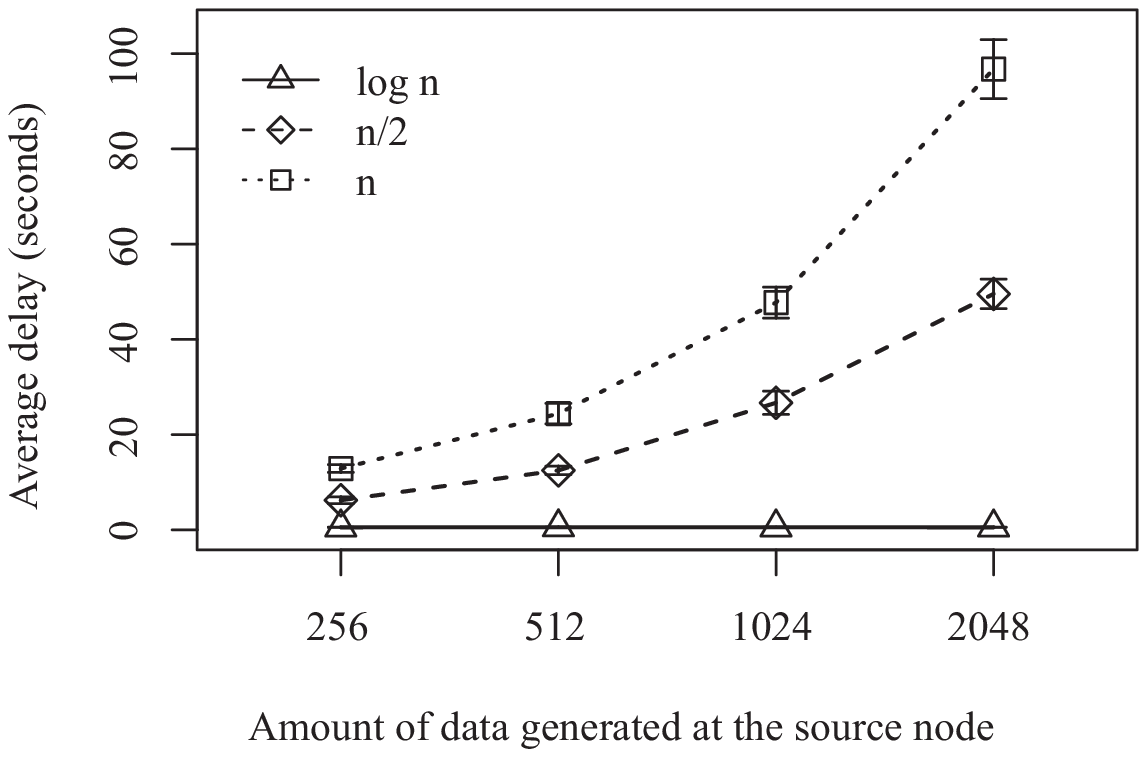}
       \label{subfig:ogkmv_STREAM_SIZE_delay_5fontes}
   }\\
   \subfigure[Varying number of nodes on the network]{
       \includegraphics[width=.45\textwidth]{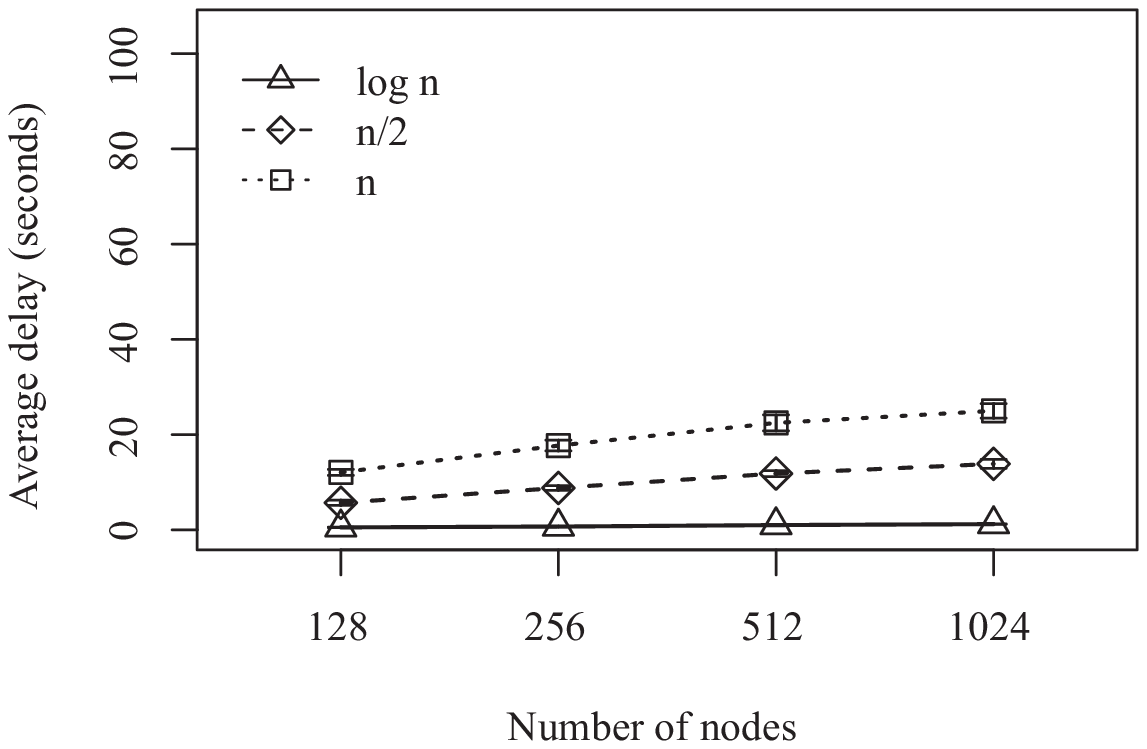}
       \label{subfig:ogkmv_NUM_NODES_delay_5fontes}
   }\\
   \subfigure[Varying number of source nodes]{
       \includegraphics[width=.45\textwidth]{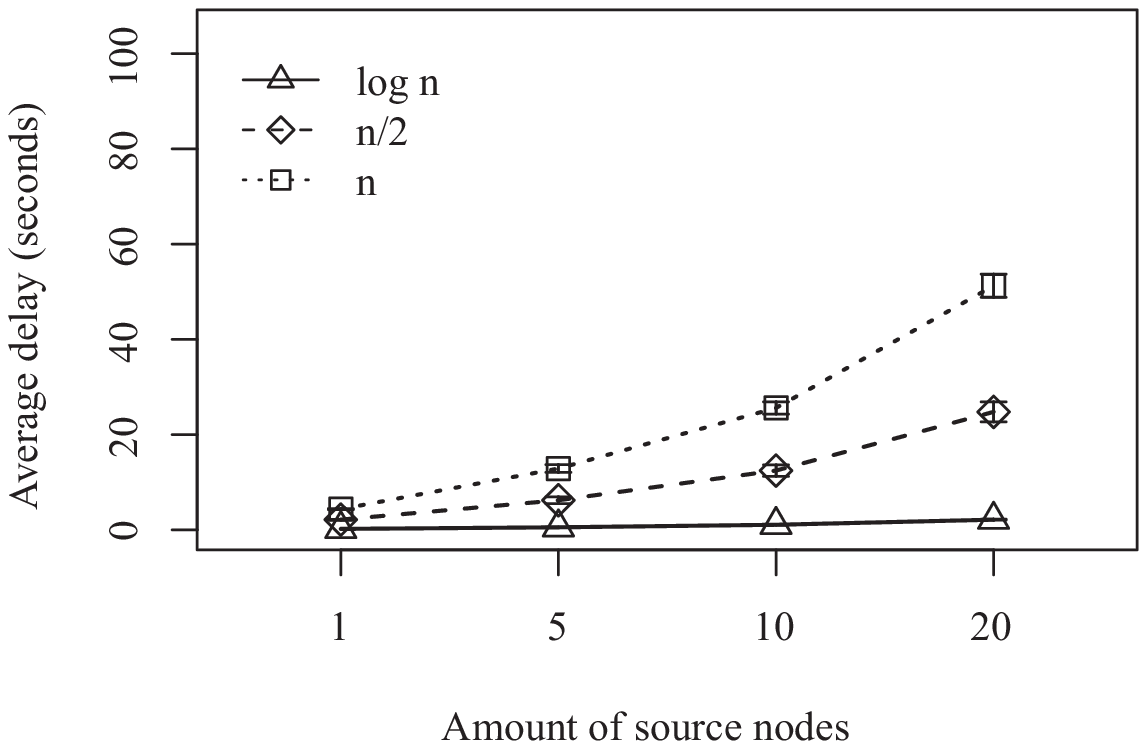}
       \label{subfig:ogkmv_SOURCE_delay}
   }
   \caption{Evaluation of packet delay}
   \label{fig:ogkmv_sensoriamento_delay}
\end{figure}

As in the energy evaluation, the first analysis is related to packet delay.
The results are presented in Fig.~\ref{fig:ogkmv_sensoriamento_delay}.

Fig.~\ref{subfig:ogkmv_STREAM_SIZE_delay_5fontes}, presents the results of varying data size $n = \{256, 512, 1024, 2048\}$, keeping fixed both the number of source nodes in $5$ and the amount of nodes in $128$.
As with energy consumption, reducing the amount of data sent by sampling reduces the delay in delivery data to the sink.
The same delay variation can be observed when generated data size increases, because more packages are sent.
Again, this variation is not observed with the $\log_{2}{n}$ sampling, since the amount of traffic on the network is very small.

The second analysis, presented in Fig.~\ref{subfig:ogkmv_NUM_NODES_delay_5fontes}, is performed varying the number of nodes in the network in $\{128, 256, 512,1024\}$ while keeping constant the amount of generated data in $n = 256$ and the number of source nodes in $5$.
In this case also is perceived the same relation observed for the energy consumption, i.e., that delay varies slightly as the number of nodes in the network is increases, since the network traffic is also almost the same.
A small increase can be observed in cases where $n$ data are sent and when $n/2$ sampling was applied.

The dependence of delay on the number of source nodes is presented in Fig.~\ref{subfig:ogkmv_SOURCE_delay}.
More once, we varied the number of source nodes in $\{1, 5, 10,20\}$, fixing the amount of data $n = 256$ and the number of nodes in $128$.
The results show the same relation observed in the energy consumption evaluation.
In this case, the delay increases considerably as the amount of source node increased, by the same reason mentioned in the previous analysis.
It is also noticeable that the delay diminishes as the amount of transmitted data is reduced.
In addition, we can also notice that under $\log_{2}{n}$ sampling, there is almost no delay variation, since the amount of transmitted packages on the network is very small.

\section{Evaluation remarks}
\label{sec:remarks}

In summary, when we analyse the data quality against the network behaviour, we have the following conclusions:
\begin{itemize}
  \item The use of MuSA reduces energy consumption and delay on the network, keeping a satisfactory data representativeness in most cases.

  \item The $n/2$ sampling is interesting even when the application requires a high level of accuracy, since both energy consumption and delay are reduced, the data are adequately preserved, and the observed errors are small even under departures from the basic model, regardless the technique.

  \item The $\log_{2}{n}$ sampling must be used only in situations where data variation is small. Otherwise, the energy consumption and delay are reduced, but the data representativeness is significantly affected.

  \item Regarding the use of PCA, robust-PCA or ICA, in most cases the three techniques presented similar results. However, robust-PCA was consistently better than the other two. In additional, the MuSA using robust-PCA has a linear time complexity $O(pn)$ instead quadratic one when MuSA uses PCA.
\end{itemize}

Finally, considering the scenarios analysed, our solution can be applied in multivariate sampling for WSNs, and the results indicate that MuSA is an efficient strategy regarding the compromise between energy consumption, delay and representativeness of the sampled data.

\section{Conclusion and future work}
\label{sec:conclusion}

WSNs have energy restrictions and the extension of their lifetime is one of most important issues in their design.
In this work, we presented an algorithm for multivariate sampling: MuSA.
MuSA uses techniques based on component analysis to classify data, making a ranking which allows selecting a subset of the most relevant data for the application.
This leads to data traffic reduction, keeping data representativeness, diminishing the energy consumption and delivery delay.

Results shown the efficiency of the proposed method regarding data representativeness.
MuSA was efficient in all evaluated scenarios with respect to both ANOVA and relative absolute error, even under departures from the ideal model (skewness and heavy-tailedness).
The best technique was consistently robust-PCA, but for a small margin over PCA and ICA.

MuSA also obtained good results with respect to network behaviour.
Sampling using MuSA resulted in considerable energy saving and packet delay reduction.
This reinforces the viability of using MuSA for performing multivariate sampling in WSNs, even in applications that need a high data precision.

This work presents a problem which arises from nodes that sample many variables concomitantly.
The volume of data in this configuration is larger than when one variable is sampled at a time.
If the volume of data is still an issue when only one variable is sampled at each epoch, the technique here presented is still valid provided (i) the process is stationary, and (ii) an adequate data storage is available.

As future work, we intend to apply the proposed method to process data along with the routing task.
Thus, not only the data from one source can be sampled, but similar data from different sources can be subjected to similar sampling strategies, resulting in even more energy efficiency.
Other aspect to be treated is the analysis of the solution in other scenarios, where data losses can affect the quality of the delivered information.

\section*{Acknowledgments}
This work is partially supported by the Brazilian National Council for Scientific and Technological Development (CNPq) and the Research Foundation of the State of Alagoas (FAPEAL).



\section*{Biography}

\begin{itemize}

\item \textbf{Andre L. L. Aquino} is Professor at Federal University of Alagoas, Brazil. He received his Ph.D. in Computer Science from the Federal University of Minas Gerais, Brazil, in 2008. His research interests include data reduction, distributed algorithms, wireless ad hoc and sensor networks, mobile and pervasive computing. In addition, he has published several papers in the area of wireless sensor networks.

\item \textbf{Orlando S. Junior} is B.S. degree in computer science from Funda\c{c}\~{a}o Pedro Leopoldo, in 2007. He received his M.S. in  informatic from Pontif\'icia Universidade Cat\'olica de Minas Gerais, in 2009. He is currently a professor at Faculdade Cenecista de Sete Lagoas and Pontif\'icia Universidade Cat\'olica de Minas Gerais. His research interests are data reduction in wireless sensor networks and software engineering.

\item \textbf{Alejandro C. Frery} graduated in Electronic and Electrical Engineering from the Universidad de Mendoza, Argentina.
His M.Sc. degree was in Applied Mathematics (Statistics) from the Instituto de Matem\'atica Pura e Aplicada (Rio de Janeiro) and his Ph.D. degree was in Applied Computing from the Instituto Nacional de Pesquisas Espaciais (S\~ao Jos\'e dos Campos, Brazil).
He is currently with the Instituto de Computa\c c\~ao, Universidade Federal de Alagoas, Macei\'o, Brazil.
His research interests are statistical computing and stochastic modelling.

\item \textbf{\'{E}dler Lins de Albuquerque} is PhD. in Development of Chemical Engineering Processes, State University of Campinas (UNICAMP) - BRAZIL, 2007. 
Professor at Federal Institute of Education, Science and Technology of Bahia. Leader of Biotechnology and Environment Research Group - BIOMA 
(\url{http://dgp.cnpq.br/buscaoperacional/detalhegrupo.jsp?grupo=0027313X6ISZ9K}). 

\item \textbf{Raquel A. F. Mini} holds a B.Sc., M.Sc., and Ph.D. in Computer Science from Federal University of Minas Gerais (UFMG), Brazil. Currently she is an Associate Professor of Computer Science at PUC Minas, Brazil.
She has worked for 10 years in the protocol design for wireless sensor networks with more than 30 papers published in this area. Her main
research areas are sensor networks, mobile computing, and ubiquitous computing.

\end{itemize}

\end{document}